\documentstyle[12pt,a4,axodraw]{article}

\input{epsf.sty}

\def\fo{\hbox{{1}\kern-.25em\hbox{l}}}

\def\slashchar#1{\setbox0=\hbox{$#1$}           
   \dimen0=\wd0                                 
   \setbox1=\hbox{/} \dimen1=\wd1               
   \ifdim\dimen0>\dimen1                        
      \rlap{\hbox to \dimen0{\hfil/\hfil}}      
      #1                                        
   \else                                        
      \rlap{\hbox to \dimen1{\hfil$#1$\hfil}}   
      /                                         
   \fi}                                         %

\def\hide#1{[hidden stuff]}

\def\beq{\begin{equation}}
\def\eeq{\end{equation}}
\def\eq{\end{equation}}
\def\to{\rightarrow}

\def\mEt{\mbox{${\hbox{$E$\kern-0.6em\lower-.1ex\hbox{/}}}_T$}\, } 

\def\bsg{\ifmmode B\to X_s\gamma\else $B\to X_s\gamma$\fi}
\def\bsll{\ifmmode B\to X_s\ell^+\ell^-\else $B\to X_s\ell^+\ell^-$\fi}
\def\bstt{\ifmmode B\to X_s\tau^+\tau^-\else $B\to X_s\tau^+\tau^-$\fi}
\def\shat{\ifmmode \hat{s}\else $\hat{s}$\fi}

\newcommand{\newc}{\newcommand}

\newc{\asusy}{\delta a^{\rm SUSY}_\mu}
\newc{\lcal}{\int {\cal L}dt}
 
\newc{\LSP}{{\chi^0_1}}
\newc{\stauR}{{\tilde \tau_R}}
\newc{\stau}{{\tilde \tau_1}}
\newc{\mstop}{m_{\tilde{t}}}
\newc{\mHpm}{m_{H^\pm}}
\newc{\gsim}{\lower.7ex\hbox{$\;\stackrel{\textstyle>}{\sim}\;$}}
\newc{\lsim}{\lower.7ex\hbox{$\;\stackrel{\textstyle<}{\sim}\;$}}
\newc{\ie}{{\it i.e.}}          
\newc{\etal}{{\it et al.}}
\newc{\eg}{{\it e.g.}}          
\newc{\kev}{\hbox{\rm\,keV}}            
\newc{\mev}{\hbox{\rm\,MeV}}            
\newc{\gev}{\hbox{\rm\,GeV}}            
\newc{\tev}{\hbox{\rm\,TeV}}
\newc{\xpb}{\hbox{\rm\, pb}}
\newc{\xfb}{\hbox{\rm\, fb}}

\newc{\mtop}{m_t}
\newc{\mbot}{m_b}
\newc{\mz}{m_Z}
\newc{\mw}{M_W}
\newc{\alphasmz}{\alpha_s(m_Z^2)}
\newc{\swsq}{\sin^2\theta_W}
\newc{\tw}{\tan\theta_W}
\newc{\cw}{\cos\theta_W}
\newc{\sw}{\sin\theta_W}
\newc{\BR}{\hbox{\rm BR}}
\newc{\zbb}{Z\to b\bar}
\newc{\Gb}{\Gamma (Z\to b\bar b)}
\newc{\Gh}{\Gamma (Z\to \hbox{\rm hadrons})}
\newc{\rbsm}{R_b^\hbox{\rm sm}}
\newc{\rbsusy}{R_b^\hbox{\rm susy}}
\newc{\drb}{\delta R_b}

\newc{\sgn}{\mbox{sgn}}
\newc{\tbeta}{\tan\beta}
\newc{\uL}{{\tilde u_L}}
\newc{\uR}{{\tilde u_R}}
\newc{\cL}{{\tilde c_L}}
\newc{\cR}{{\tilde c_R}}
\newc{\tL}{{\tilde t_L}}
\newc{\tR}{{\tilde t_R}}
\newc{\dL}{{\tilde d_L}}
\newc{\dR}{{\tilde d_R}}
\newc{\sL}{{\tilde s_L}}
\newc{\sR}{{\tilde s_R}}
\newc{\bL}{{\tilde b_L}}
\newc{\bR}{{\tilde b_R}}
\newc{\eL}{{\tilde e_L}}
\newc{\eR}{{\tilde e_R}}
\newc{\mhp}{m_{H^\pm}}
\newc{\mhalf}{m_{1/2}}
\newc{\emt}{{e/\mu /\tau}}

\newc{\lR}{\tilde{l}_R}
\newc{\lL}{\tilde{l}_L}
\newc{\nL}{\tilde{\nu}_L}
\newc{\na}{\chi^0_1}
\newc{\nb}{\chi^0_2}
\newc{\nc}{\chi^0_3}
\newc{\nd}{\chi^0_4}
\newc{\ca}{\chi^{\pm}_1}
\newc{\cb}{\chi^{\pm}_2}
\newc{\camp}{\chi^\mp_1}
\newc{\cbmp}{\chi^\mp_1}
\newc{\capos}{\chi^{+}_1}
\newc{\caneg}{\chi^{-}_1}
\newc{\phit}{\phi_t}
\newc{\phib}{\varphi_b}
\newc{\phiew}{\phi_{ew}}
\newc{\htz}{h^0_t}
\newc{\hbz}{h^0_b}
\newc{\hewz}{h^0_{ew}}
\newc{\hsmz}{h^0_{sm}}
\newc{\huz}{h^0_u}
\newc{\hsusyz}{h^0_{susy}}

\newcommand{\drawsquare}[2]{\hbox{%
\rule{#2pt}{#1pt}\hskip-#2pt
\rule{#1pt}{#2pt}\hskip-#1pt
\rule[#1pt]{#1pt}{#2pt}}\rule[#1pt]{#2pt}{#2pt}\hskip-#2pt
\rule{#2pt}{#1pt}}

\newc{\Dal}{\drawsquare{7}{0.6}}

\def\dofig#1#2{\epsfxsize=#1\centerline{\epsfbox{#2}}}
\def\dofigs#1#2#3{\centerline{\epsfxsize=#1\epsfbox{#2}%
   \hfil\epsfxsize=#1\epsfbox{#3}}}

\def\beq{\begin{equation}}
\def\eeq{\end{equation}}
\def\bea{\begin{eqnarray}}
\def\eea{\end{eqnarray}}
\catcode`@=11
\long\def\@caption#1[#2]#3{\par\addcontentsline{\csname
  ext@#1\endcsname}{#1}{\protect\numberline{\csname
  the#1\endcsname}{\ignorespaces #2}}\begingroup
    \small
    \@parboxrestore
    \@makecaption{\csname fnum@#1\endcsname}{\ignorespaces #3}\par
  \endgroup}
\catcode`@=12

\begin{document}
\begin{titlepage}

\begin{flushright}
hep-ph/0108006 \\
LBNL-48649 \\
\end{flushright}

\huge
\bigskip
\bigskip
\begin{center}
{\Large\bf
Superlight gravitinos in electron-photon collisions}
\end{center}

\large

\vspace{.15in}
\begin{center}

Shrihari Gopalakrishna and James Wells

\small

\vspace{.1in}
{\it Physics Department, 
               University of California, Davis, CA 95616}, and\\
{\it Theoretical Physics Group, Lawrence Berkeley National 
 Laboratory, Berkeley, CA 94720}\\

\end{center}
 
\vspace{0.15in}
 
\begin{abstract}

Motivated by recent studies of supersymmetry in higher-dimensional
spaces, we discuss the experimental signatures of a superlight
gravitino (${\rm mass} \leq 10^{-3}\, {\rm eV}$).
We concentrate on the
process $e^- \gamma \rightarrow \tilde e_R\tilde G$ as a probe of
supersymmetry, where a single heavy superpartner and a superlight
gravitino are produced. The fact that there is only one heavy
superpartner in the final state in this process would require a lower
center-of-mass energy for on-shell production compared to conventional pair
production. For instance, for a 500~GeV machine, we find that a positive signal will be found if the supersymmetry breaking scale is less than about 2~TeV. If no positive signal is found, this process puts a
bound on the supersymmetry breaking scale.

\end{abstract}

\medskip

\begin{flushleft}
July 2001
\end{flushleft}

\end{titlepage}

\baselineskip=18pt


The gravitino is a spin $3/2$ superpartner to the graviton.
The massless
gravitino has two spin degrees of freedom, the $+3/2$ and the $-3/2$
helicity components.
When supersymmetry is spontaneously broken, the gravitino acquires a
mass, denoted by $m_{3/2}$. The precise value of $m_{3/2}$ is model
dependent. One possible way in which supersymmetry can be 
broken spontaneously is by the $F$ component vacuum expectation value 
of a hidden sector superfield,
$Z=(z,\chi,F)$.  ($z$ is
a scalar, $\chi$ is a spin $1/2$ fermion and $F$ is an auxiliary
field.) The massive gravitino has four spin degrees of freedom, and
the two additional (longitudinal) ones are supplied by the goldstino,
$\chi$. In this case, the mass of the gravitino is 
\bea
m_{3/2} & = & \frac{\langle F\rangle}{\sqrt{3}M_{\rm pl}},  \ \ \  
M_{\rm pl} = (8\pi G_{N})^{-1/2} \approx 2.4 \times 10^{18}\gev.
\eea

The scale of supersymmetry breaking, $\langle F\rangle$, and hence the
mass of the gravitino, are presently not known. It is instructive to
note what the different schemes of supersymmetry-breaking imply for
$m_{3/2}$. Weak-scale supersymmetry breaking
and some
extra dimensions theories
can give superlight gravitinos~\cite{Gherghetta:2000qt,Gherghetta:2001kr},
$m_{3/2} \lsim 10^{-3}\, {\rm eV}$, $\sqrt{F} \lsim 10^3\gev$; Gauge
Mediation, $m_{3/2} \approx 1~\, {\rm eV}-10\kev$, $\sqrt{F} \approx
10^5-10^7\gev$; Generic Gravity Mediation, $m_{3/2} \approx
10\kev-10\tev$, $\sqrt{F} \approx 10^7-10^{12}\gev$; Anomaly Mediation,
$m_{3/2} > 10\tev$, $\sqrt{F} > 10^{12}\gev$.

There are collider, cosmological and astrophysical bounds on the mass
of the gravitino. The current LEP bound due to non-observation of any
deviation from the Standard Model is $m_{3/2} \gsim
10^{-5}\, {\rm eV}$~\cite{Brignole:1998sk}. The constraints from the cooling of
stars excludes the region $10^{-2}\, {\rm eV} \lsim m_{3/2} \lsim
10^{2}\, {\rm eV}$~\cite{Nowakowski:1995ag}. There is thus a window 
$10^{-5}\, {\rm eV}
\lsim m_{3/2} \lsim 10^{-2}\, {\rm eV}$ for the mass of the gravitino and we
focus on this range of masses and analyze its implications for
electron-photon collisions. As we will discuss in the following
section, the production cross-section is inversely proportional to the
square of $m_{3/2}$ (equivalently, $\sigma \sim 1/F^2$)
and thus superlight gravitinos become interesting
at next generation colliders.


Owing to the Goldstone equivalence theorem, at energies much bigger
than $m_{3/2}$, the production cross-section of the longitudinal
helicity components ($+1/2$ and $-1/2$) is approximately equal to the
production cross-section of the goldstino, $\chi$, that  was eaten. The
correction is of order $m_{3/2}/E$ where $E$ is the energy of the
process. To get the amplitudes for the production of the longitudinal
helicity components of the gravitino, it is sufficient to replace the
gravitino field $\psi_\mu$ by,
\bea
\psi_\mu \sim i\sqrt{\frac{2}{3}}\frac{1}{m_{3/2}}\partial_\mu\tilde G,
\eea
where $\tilde G$ is the goldstino.
We focus on the goldstino production ($+1/2$ and $-1/2$ components)
since the production amplitude is proportional to $1/F$ 
which is significantly stronger than the
$1/M_{\rm pl}$ coupling for
the $+3/2$ and $-3/2$ helicity components. This can be seen from the
Lagrangian given below.

We focus on a scenario in which the gravitino is the superlight LSP, $\tilde
e_R$ the NLSP, and all other superpartners are much heavier. This is
for instance the case in the
model~\cite{Gherghetta:2000qt,Gherghetta:2001kr} constructed in
the Randall-Sundrum
scenario~\cite{Randall:1999ee}. In this framework,
we live on a 3-dimensional brane in AdS space, an idea 
motivated by the need to generate two hierarchically
different scales, $M_{\rm pl}$ and $M_{\rm weak}$.
The hierarchically different masses of the gravitino and the gaugino 
are determined directly~\cite{Gherghetta:2001kr} 
from the poles in the propagators on the brane.
This computation gives a
superlight gravitino and gauginos with TeV scale masses. The scalars also
get
TeV scale masses due to radiative corrections. A more intuitive 
reason~\cite{Gherghetta:2001kr} for the superlight gravitino appeals to the
conjectured AdS/CFT correspondence. In the CFT picture, Supersymmetry
breaking is
due
to a strongly coupled sector charged under the Standard Model gauge group,
which
results in a tree-level gaugino mass of TeV scale. The gravitino coupling
to the CFT has a $1/M_{\rm pl}$ suppression and hence gets a mass of order
$\tev^2/M_{\rm pl}$.

For the Goldstino production in the process $e^- \gamma \rightarrow
\tilde e_R\tilde G$, it is sufficient to work with an effective global
supersymmetric lagrangian. The relevant terms in this effective
lagrangian are~\cite{Brignole:1998sk,Gherghetta:1997fm,Brignole:1997pe}
\bea
L &\supset& - \frac{\tilde m^2}{|M|}\left[\frac{M}{F}
(\tilde e_R \bar e P_L \tilde G) 
+ \frac{M^*}{F^*}(\tilde e^*_R \bar{\tilde G} P_R e)\right] -\sqrt{2}ig
(\tilde e^*_R \bar{\lambda} P_R e - \tilde e_R \bar e P_L \lambda) \nonumber \\
&-& \frac{1}{4\sqrt{2}} \bar{\tilde G}[\gamma^{\mu},\gamma^{\nu}]\lambda \left[{\rm Re}\left(\frac{M}{F}\right) F_{\mu\nu} - {\rm Im}\left(\frac{M}{F}\right){\tilde F}_{\mu\nu} \right]
+ igA_\mu(\tilde e^*_R \partial^\mu \tilde e_R - \partial^\mu \tilde e^*_R \tilde e_R)
\label{Lmatcoup}
\eea
where ${\tilde{m}}^2$ is the $\tilde e_R$ mass squared, $\tilde G$ 
the Goldstino, $e$ the electron, and $M$ the mass of the gaugino $\lambda$.
The gaugino is included above, since it contributes to the effective 
four point vertex, as shown in the Feynman diagrams below:


\begin{center}
\begin{picture}(325,100)(0,0)
\Photon(0,20)(30,50){3}{5}
\Text(15,30)[lt]{$\gamma$}
\ArrowLine(30,50)(70,50)
\Text(50,55)[b]{$e^-$}
\ArrowLine(0,80)(30,50)
\Text(15,70)[lb]{$e^-$}
\Line(70,50)(100,20)
\Text(85,30)[rt]{$\tilde G$}
\DashArrowLine(70,50)(100,80){2}
\Text(87,70)[rb]{${\tilde e}_R^-$}
\Photon(130,0)(160,30){3}{5}
\Text(145,12)[lt]{$\gamma$}
\ArrowLine(130,100)(160,70)
\Text(145,88)[lb]{$e^-$}
\Line(160,70)(190,100)
\Text(175,85)[rb]{$\tilde G$}
\DashArrowLine(160,70)(160,30){2}
\Text(163,50)[l]{${\tilde e}_R^-$}
\DashArrowLine(160,30)(190,0){2}
\Text(175,15)[rt]{${\tilde e}_R^-$}
\Photon(220,15)(250,45){3}{5}
\Text(235,27)[lt]{$\gamma$}
\ArrowLine(220,85)(250,55)
\Text(235,73)[lb]{$e^-$}
\DashArrowLine(250,55)(280,85){2}
\Text(265,70)[rb]{${\tilde e}_R^-$}
\SetWidth{1.5}
\Line(250,55)(250,45)
\SetWidth{0.5}
\Text(253,50)[l]{$\lambda$}
\Line(250,45)(280,15)
\Text(265,30)[rt]{$\tilde G$}
\end{picture}
\end{center}

For producing $\eR$ the incoming $e^-$ must be right handed. 
The photon helicities can be $+$ or $-$.   The amplitudes
for these two photon helicities in the center-of-mass frame are
\bea
{\cal M}_+ & = &\frac{e \tilde m^2}{F} \sqrt{\frac{2 { p}}{E}}
\left[\left(1+\frac{4E^2}{\tilde m^2}\right)\sqrt{\frac{1-\cos\theta}{2}} - 
\frac{\sin\theta\sqrt{\frac{1+\cos\theta}{2}}}{(1+\cos\theta)
+{\tilde m^2}/{2E{p}}}\right],  \\
{\cal M}_- & = & \frac{e \tilde m^2}{F} \sqrt{\frac{2 {p}}{E}}
\left[\frac{-\sin\theta\sqrt{\frac{1+\cos\theta}{2}}}{(1+\cos\theta)
+{\tilde m^2}/{2E{p}}}
\right],
\eea
where $\theta$ 
is the angle made by the $\eR$ with respect to the incoming $e^-$,
$E$ is the incoming $e^-$ beam energy, $p\equiv |{\bf p}|$ is the 3-momentum 
magnitude of the $\eR$, $\tilde m$ is the $\eR$ mass,
and $e=\sqrt{4\pi\alpha}$.
The cross-sections in the center-of-mass frame are,
\bea
\frac{d~\sigma_\pm}{d\cos\theta} = \frac{(1-{\tilde m^2}/
{4E^2})}{128\pi E^2}|{\cal M}_\pm|^2.
\eea

We are concentrating on the process $e^- \gamma \rightarrow \eR \tilde G$ as
a probe of supersymmetry, since on-shell production requires the
center-of-mass energy of the collision only
be greater than the mass of
one heavy superpartner ($\eR$) as opposed to twice this in
conventional pair production of heavy superpartners.
The $\eR^-$ produced decays promptly to an $e^-$ and a $\tilde G$ (the
$\tilde G$ escapes the detector unseen). Thus the experimental
signature in the detector is  $e^-$ +\slashchar{E}.  This is similar
to the signal of $\tilde e\chi^0_1$ production in $e^-\gamma$
collisions~\cite{Kon:1992vg,Barger:1998qu,Kiers:1996ux}.

The background is due to the Standard Model processes: $e^- \gamma
\rightarrow e^- Z \rightarrow e^- \nu \bar{\nu}$ and $e^- \gamma
\rightarrow W^- \nu_e  \rightarrow e^- \nu_e \bar{\nu_e}$. The
background can be reduced significantly by placing appropriate cuts
and by using polarized beams as will be discussed later. Helicity
amplitudes for the background are given in \cite{Denner:1993qf} and
\cite{Denner:1993vg}.

\begin{figure}[tpb]
\dofig{3.75in}{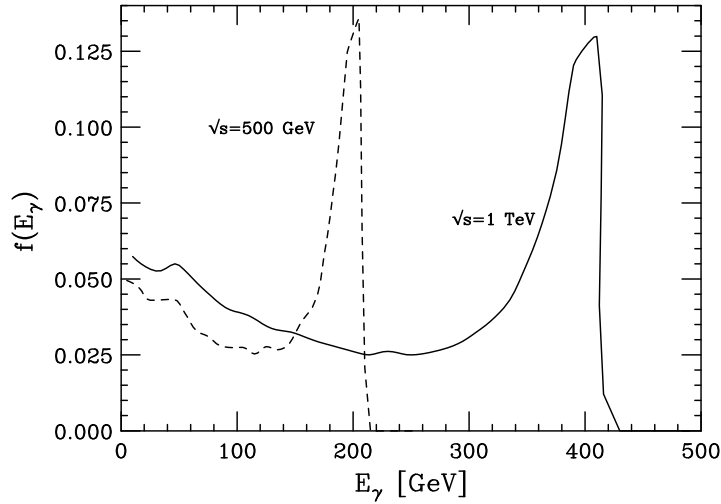}
\caption{The Compton back scattered photon spectra obtained by scattering 
1~eV photons with polarization $-1.0$ on (i) 250~GeV electrons and (ii)
500~GeV electrons. The electron beams have polarization $+0.8$.
\label{gspect}}
\end{figure}

The high energy photons needed for the $e^-\gamma$ collisions
are produced by Compton back scattering a low-energy laser
beam from a second electron beam. The Compton back scattered photons
have an energy spread as shown in Fig.~\ref{gspect}, with data from the package Pandora~\cite{Peskin:1999hh}.

In the following, we present results for two machine designs: 
\begin{itemize}
\item ``500~GeV collider'': Collision of 250~GeV ($+0.8$ polarized) electrons
with the photons whose spectrum is shown in Fig.~\ref{gspect}. 
We denote this as a 500~GeV machine since the maximum
electron-photon center-of-mass 
energy is close to, but somewhat less than 500~GeV.
That is, the 250~GeV electron beam collides against backscattered photons
with maximum beam energy slightly less than 250~GeV, yielding a maximum
center of mass energy for $e^-\gamma$ collisions slightly less than
500~GeV.
\item ``1TeV collider'': Collision of 500~GeV ($+0.8$ polarized) electrons
with the photons whose spectrum is shown in Fig.~\ref{gspect},
with center-of-mass energy slightly less than 1~TeV.
\end{itemize}

\begin{figure}[tpb]
\dofigs{3.5in}{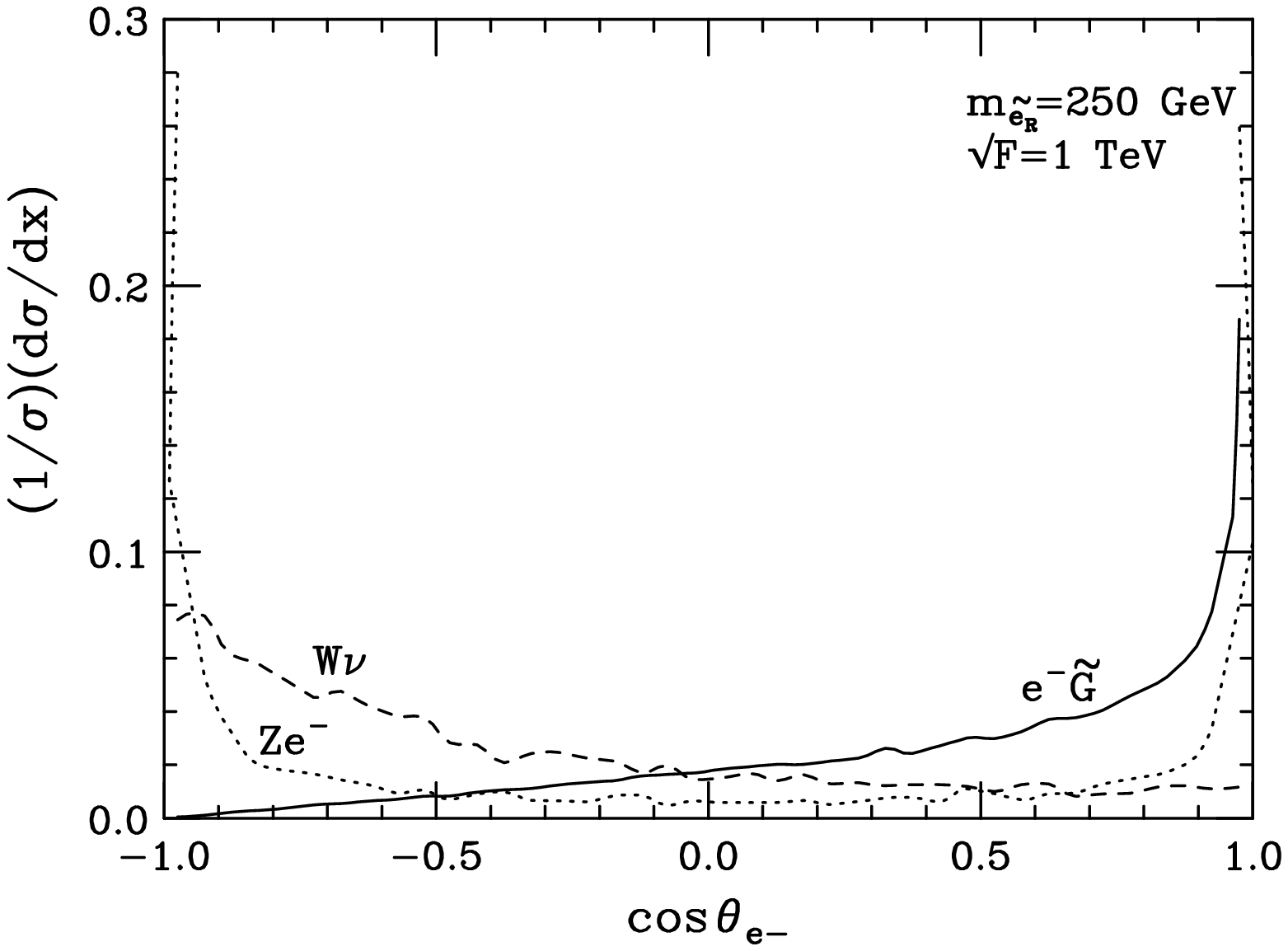}{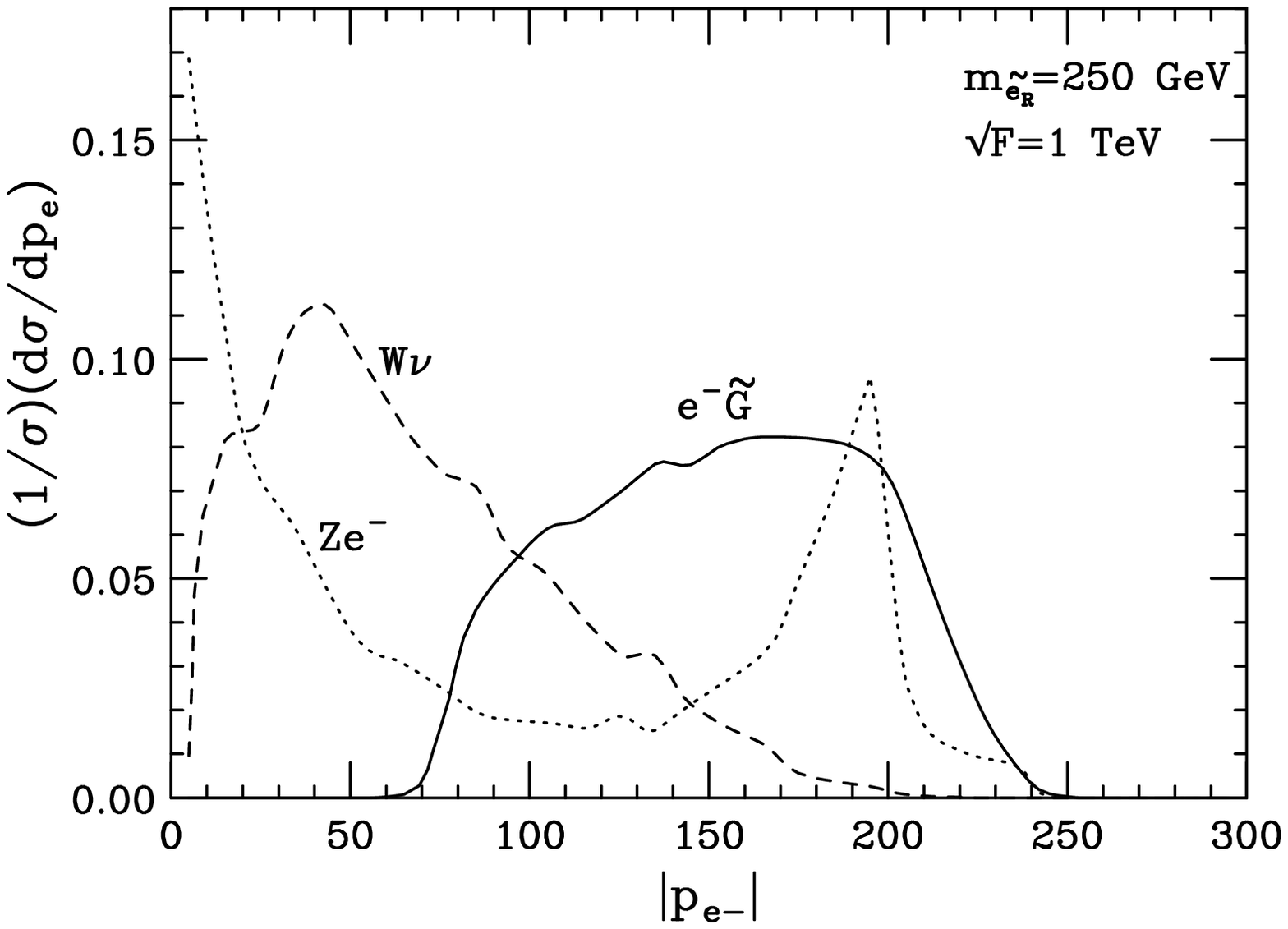}
\caption{The differential signal and background cross-sections as a function 
of the outgoing $e^-$ $cos(\theta)$ and $|{\bf p}|$. The signal cross-section 
is shown for a representative choice of ($\tilde m=250\gev$, $\sqrt{F}=1\tev$).
\label{diffcs}}
\end{figure}

We used the package Pandora~\cite{Peskin:1999hh} 
to obtain the results in this section with
the unpolarized cross-sections verified with Pythia and our own
Monte Carlo code.
Fig.~\ref{diffcs} shows the differential cross-section as a function of
$\cos\theta$ and $|{\bf p}|$ of the outgoing $e^-$ for the case of a
500~GeV collider. The cross-section shown is for the production of the $e^-$
after factoring in the branching fraction to $e^-$. $\theta$ is
measured from the direction of the incident $e^-$ beam. The background
is reduced by applying a theta cut to accept events in the
range $10^\circ < \theta < 125^\circ$, and a momentum cut is employed 
to accept events with $|{\bf
p}_e|$ $> 100\gev$ for the case of $\tilde m=250\gev$. A different
momentum cut is found to be optimal for other choices of $\tilde m$.

\begin{figure}[tpb]
\dofig{3.75in}{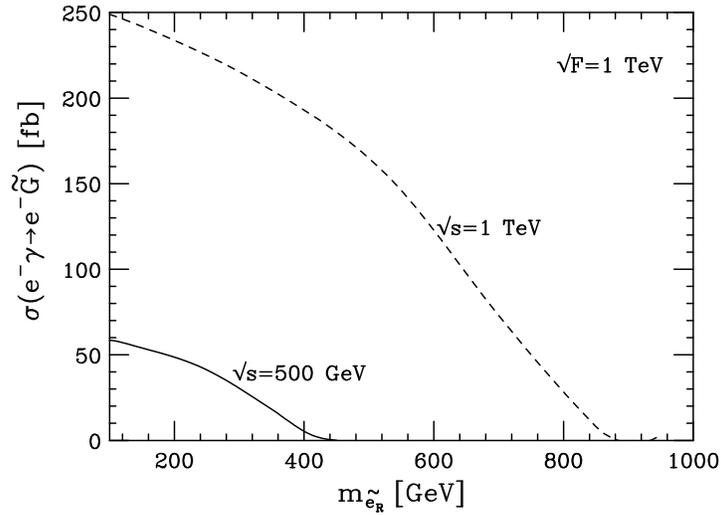}
\caption{The dependence of the $\eR$ production cross-section 
on $\tilde m$ for a 500~GeV collider and a 1~TeV collider.
\label{csvsm}}
\end{figure}

Fig.~\ref{csvsm} shows the dependence of the cross-section on the mass
of the selectron. It should be noted that the selectron production
cross-section scales as $1/(\sqrt{F})^2$ as can be inferred from
eq.~\ref{Lmatcoup}.

\begin{figure}[tpb]
\dofigs{3.5in}{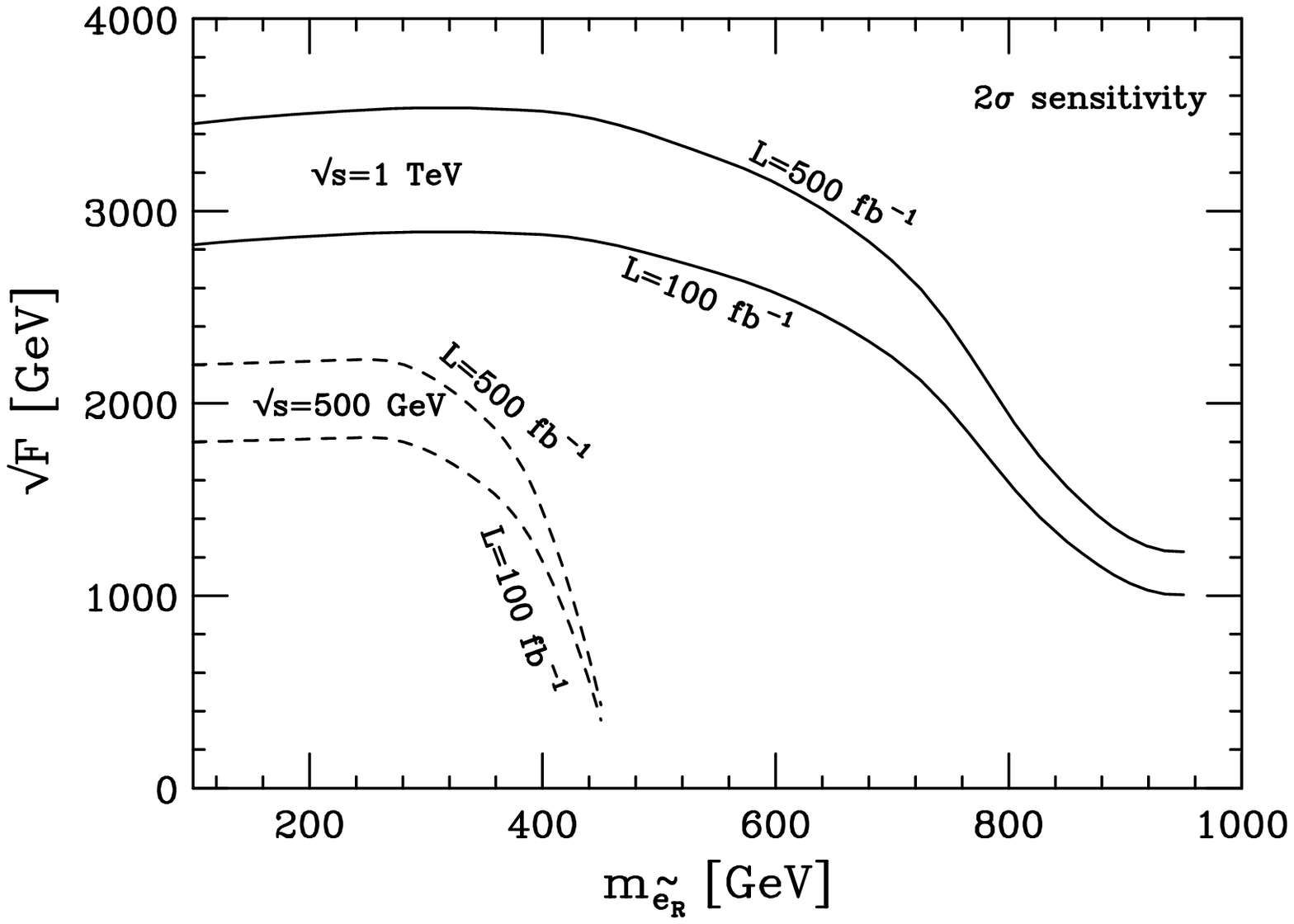}{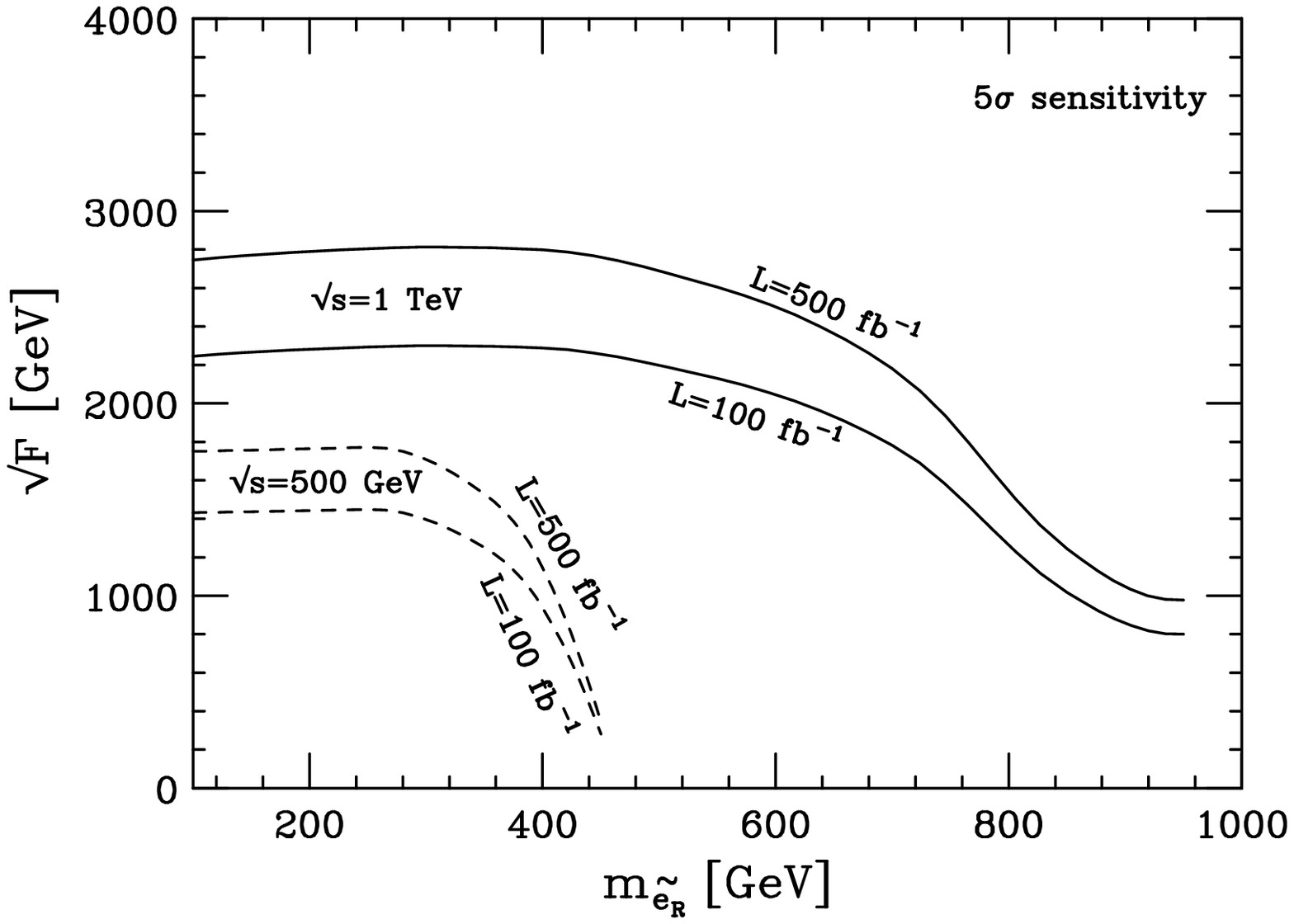}
\caption{$2 \sigma$ (left) and $5 \sigma$ (right) reach for a 
500~GeV collider and a 1~TeV collider 
with luminosities of $100\xfb^{-1}$ and $500\xfb^{-1}$.  
\label{reachfig}}
\end{figure}

Fig.~\ref{reachfig} shows the reach of a 500~GeV collider and a 1~TeV collider 
for luminosities of $100\xfb^{-1}$ and $500\xfb^{-1}$. Momentum and angle cuts
were placed on the outgoing electron to suppress the background and
enhance the reach. The 500~GeV collider curve shows the reach after
making the theta cut $10^\circ < \theta < 125^\circ$ and the $|{\bf
p}|$ cuts: ($\tilde m$(GeV), $|{\bf p}|$(GeV)) $\rightarrow$ (100,50),
(150,50), (250,100), (350,125), (400,150), (450,150). The 1~TeV collider
curve is with the same theta cut, $10^\circ < \theta < 125^\circ$, and
a momentum cut at $100\gev$. The plot on the left shows a $2 \sigma$
exclusion contour; if a positive signal is not found, the region below
the line would be excluded. In the plot on the right, the region below
the lines will result in a $5 \sigma$ excess over background.

We have presented the mode $e^- \gamma \rightarrow \tilde e_R\tilde G$
as a probe of supersymmetry and as a means of gleaning insight into
the scale of supersymmetry breaking. The advantage of this mode over
conventional pair production is that this requires just the mass of
one heavy superpartner in the center of mass before it is produced
on-shell.

Polarized beams can offer tremendous advantages for suppressing $W^-\nu$ and $Z e^-$ backgrounds. This results in the excellent reach for probing the supersymmetric breaking scale $\sqrt F$.
Our studies indicate that the reach for a 500~GeV (1~TeV) machine design is nearly 2~TeV (3~TeV) with $500\xfb^{-1}$ of integrated luminosity.

Another useful way to search for light gravitinos is through
$e^+e^-\to \tilde G\tilde G\gamma$ production.  A finite cross-section
is still possible even if all the other superpartners are extremely
heavy.
The current LEP lower bound on $\sqrt{F}$ from this process is around
$200\gev$~\cite{Brignole:1998sk} ($\sqrt{s} = 183\gev$ with $60\xpb^{-1}$
of data). Extrapolating this to an $e^+e^-$ linear collider with $100\xfb^{-1}$
gives a $\sqrt{F}$ reach of $\sim 750\gev$ for a 500~GeV collider 
and $\sim 1\tev$ for a 1~TeV collider. This estimate is for 
an unpolarized $e^+e^-$
machine and can be improved upon by using polarized beams to reduce
the background. The reach of the upgraded Tevatron is $\sim 410\gev$
and that of the LHC $\sim 1.6\tev$~\cite{Brignole:1998me}.
Although our signal of single gravitino production is sensitive to somewhat
higher values of $\sqrt{F}$, the $\tilde G\tilde G\gamma$ signal does not
require another accessible superpartner and should be considered
complementary.

This study was done with a realistic Compton backscattered photon
energy spectrum. Only primary backscatters were considered and
secondary scatters ignored. Including secondary scatters gives a
higher luminosity in the low-energy tail of the photon energy spectrum
and its effect on the reach should be minimal.

\noindent
{\it Acknowledgements:}
SG wishes to acknowledge Lawrence Livermore Lab for financial support,
and M.~Peskin for help implementing the helicity amplitudes
into Pandora.  We also wish to thank 
T.~Gherghetta, J.~Gronberg, J.~Gunion, A.~Konechny, S.~Mrenna and B.~Murakami for valuable discussions.



\begin{thebibliography}{20}

\bibitem{Gherghetta:2000qt}
T.~Gherghetta and A.~Pomarol,
``Bulk fields and supersymmetry in a slice of AdS,''
hep-ph/0003129.

\bibitem{Gherghetta:2001kr}
T.~Gherghetta and A.~Pomarol,
``A warped supersymmetric standard model,''
Nucl.\ Phys.\ B {\bf 602}, 3 (2001)
[hep-ph/0012378].

\bibitem{Brignole:1998sk}
A.~Brignole, F.~Feruglio and F.~Zwirner,
``Signals of a superlight gravitino at $e^+ e^-$ 
colliders when the other superparticles are heavy,''
Nucl.\ Phys.\  {\bf B516}, 13 (1998)
[hep-ph/9711516].

\bibitem{Nowakowski:1995ag}
M.~Nowakowski and S.~D.~Rindani,
``Astrophysical limits on gravitino mass,''
Phys.\ Lett.\ B {\bf 348}, 115 (1995)
[hep-ph/9410262].

\bibitem{Randall:1999ee}
L.~Randall and R.~Sundrum,
``A large mass hierarchy from a small extra dimension,''
Phys.\ Rev.\ Lett.\  {\bf 83}, 3370 (1999)
[hep-ph/9905221].

\bibitem{Gherghetta:1997fm}
T.~Gherghetta,
``Goldstino decoupling in spontaneously broken supergravity theories,''
Nucl.\ Phys.\  {\bf B485}, 25 (1997)
[hep-ph/9607448].

\bibitem{Brignole:1997pe}
A.~Brignole, F.~Feruglio and F.~Zwirner,
``On the effective interactions of a light gravitino with matter fermions,''
JHEP {\bf 9711}, 001 (1997)
[hep-th/9709111].

\bibitem{Kon:1992vg}
T.~Kon and A.~Goto,
``Single selectron production at TeV $e \gamma$ colliders,''
Phys.\ Lett.\  {\bf B295}, 324 (1992).

\bibitem{Barger:1998qu}
V.~Barger, T.~Han and J.~Kelly,
``Sparticle production in electron photon collisions,''
Phys.\ Lett.\  {\bf B419}, 233 (1998)
[hep-ph/9709366].

\bibitem{Kiers:1996ux}
K.~Kiers, J.~N.~Ng and G.~Wu,
``Supersymmetric Signatures at an $e\gamma$ Collider,''
Phys.\ Lett.\  {\bf B381}, 177 (1996)
[hep-ph/9604338].

\bibitem{Denner:1993qf}
A.~Denner and S.~Dittmaier,
``Electroweak radiative corrections to $e^- \gamma \to e^- Z$,''
Nucl.\ Phys.\ B {\bf 398}, 265 (1993).

\bibitem{Denner:1993vg}
A.~Denner and S.~Dittmaier,
``Electroweak radiative corrections to $e^- \gamma \to W^- \nu_e$,''
Nucl.\ Phys.\ B {\bf 398}, 239 (1993).

\bibitem{Peskin:1999hh}
M.~E.~Peskin,
``Pandora: An object-oriented event generator for linear collider  physics,''
hep-ph/9910519.

\bibitem{Brignole:1998me}
A.~Brignole, F.~Feruglio, M.~L.~Mangano and F.~Zwirner,
``Signals of a superlight gravitino at hadron colliders when 
the other superparticles are heavy,''
Nucl.\ Phys.\  {\bf B526}, 136 (1998)
[hep-ph/9801329].

\end{thebibliography}
\end{document}